# The Growth of Ordered Dusty Structures in Glow Discharge Plasma.

A. Khakhaev, S. Podriadtchikov
*Petrozavodsk State University, Russia*

1. **Phenomenon of self-organization of ordered dusty structures in plasma as an experimental fact.**

    The presence of condensed disperse phase (CDP) and electrical fields in different kinds of plasma leads to the forming of space-limited dusty clouds. These structures may exist like a self-organized structure of all kinds.
In experimental investigations the following dusty structures were observed:
1   ring-shaped structures
2   tear-shaped structures
3   spherical structures
4   coiled configuration structures
5   "tornado"-like structures

    The oscillations' absence of dusty particles in structure near their equilibrium position, including oscillations, which appear due to excitations of density waves, was registered. These waves are self-excited in the dusty structure under certain conditions or may be initiated by external influence.

    It was discovered that there are condition fields in plasma, in which the rate of self-organization varies but at the same time an average number of particles involved in the structure does not change. As a result, the existence of different condition phases of ordered structure was proved. Some kinds of structures are presented in fig.1

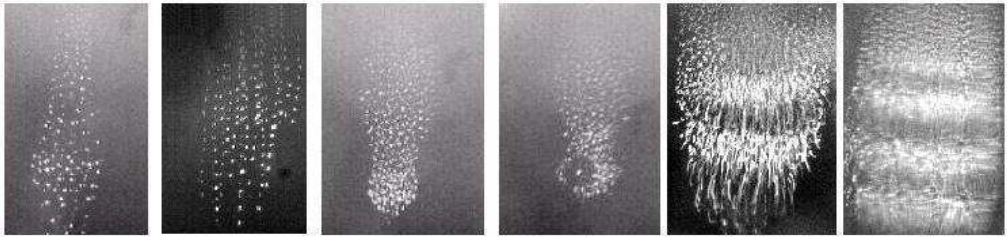

Fig.1. Different kinds of dusty structures.

2. **Some problematic aspects of ordered structures research in complex plasma.**

    Interaction of dusty macroparticles with plasma components may lead to the changes in plasma conditions, its atomic-molecular composition and matter modification, macroparticles surface behavior in particular. Therefore, the properties of complex plasma may be sufficiently different compared to the properties of plasma without CDP.

    The presence of ordered structures (located, for example, in positive column of a glow discharge) changes the distribution of a spatial charge (a form of spatial fields), local values of conductivity, and dispersive properties of plasma in structure localization area.

    The theoretical description of complex plasma properties' formation is only being started, because of shortage of integrated experimental results, on which mentioned above theory may be based.

    Among the unsolved theoretical and practical aspects we can list the following:

    1. It is still not clear how many particles can take part in the formation of ordered structure under certain conditions

    2. We can't define how many and what kind of structures self-organize in a plasma-CDP-

force-field-configuration system

3. What structures and phase conditions will work in the mentioned above system, how many macroparticles will be involved.

The attempts at a theoretical prediction of existence areas of various phase conditions in described above system [1] are based on operation with the parameters, which can not be considered independent:

nonideality parameter

$$\Gamma = \frac{Z^2 e^2}{akT_g}$$

, where $Ze$, $T_g$ - are particle charge and temperature, $a$- interparticle distance, $k$-Boltzmann constant;

modified nonideality parameter

$$\Gamma^* = \Gamma(1+\chi+\frac{\chi^2}{2})\exp(-\chi)$$

where $\chi=a/r_D$, $r_D$- the Debye radius.
It is accepted that the condition of crystallization is $\Gamma^*>106$ or more detail it is presented in fig.2

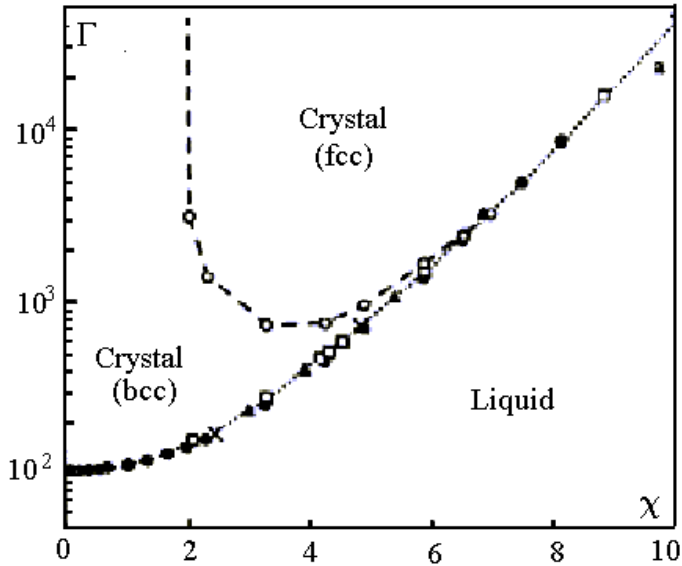

Fig 2. Phase transition criteria.

The models' verification requires a direct measurement of $a$, $Z$, $T_g$ and plasma parameters, which determine $r_D$. The discussion of some questions and the current researches were published. [1,2,3].

Theoretical researches and experimental experience allowed finding out perspective approaches to data acquisition, necessary for the development of the ideas of self-organization end ordered structure existence in the complex plasma. These approaches include the improvement of methods and experimental equipment for obtainment of new data on space-time macroparticles' distribution during the process of ordered structures' formation, existence and destruction. For instance:

- The usage of optic-spectral methods of plasma local characteristics' analysis with and without dusty structure.
- The integrated automation of acquisition and processing of tomography, optic-spectral and electrical data in order to provide its objectivity, statistical reliability, accuracy, representation and storing.

3. The experimental setup for the research of complex plasma with ordered structures, used in the Research Education Center of PetrSU (REC-013) is presented schematically on fig 3.

It provides the connection between the vacuum installation and various plasma generators using the RF-discharge, the DC-discharge, etc. as well as necessary vacuum conditions (up to P=10-8 Pa, pressure during experiment P>10 Pa), carrying out experiments in automatic mode with simultaneous registration of a photo image or creation of videofilms of macroparticles with CDP.
Using a cylindrical lens laser light is formed into the laser "knife" and so scattering into macroparticles light is captured by videocamera.

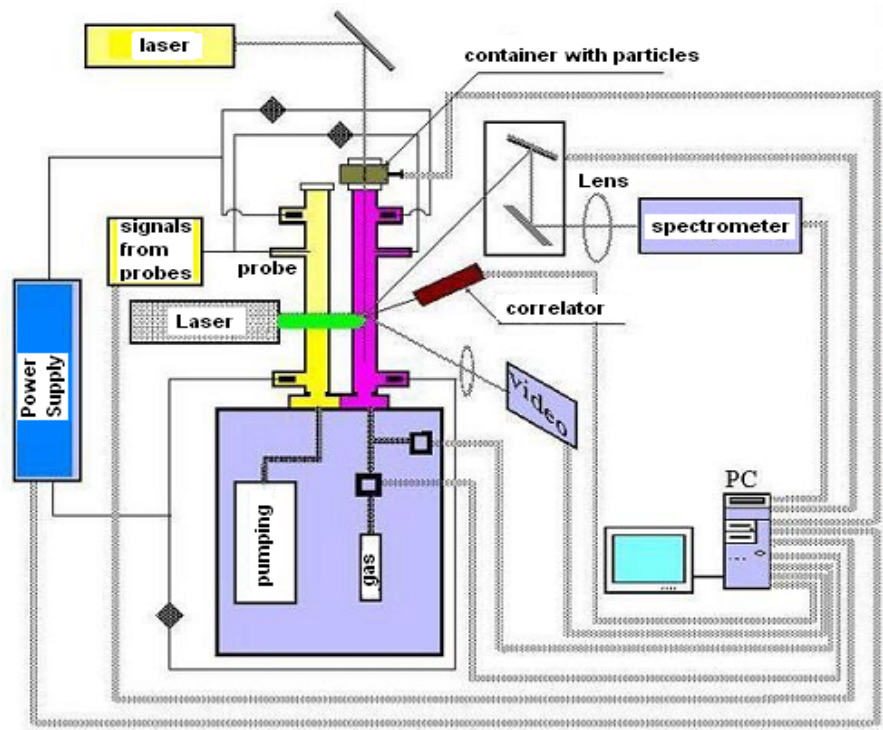

Fig. 3. The experimental setup.

Using a cylindrical lens laser light is formed into the laser "knife" and so scattering into macroparticles light is captured by videocamera. Furthermore, values of discharge current, voltage and pressure, characteristics of spectral radiation from dusty structure localization area are also registered in automatic mode. Image projection of dusty structures in discharge is changed manually. Along with the mentioned above, the registration system includes correlator, which measures scattering light fluctuation (as well as the dynamics of macroparticles movement) and calculates the current phase condition of macroparticle structure. Register systems and experiment procedures are managed with the G-language LabView software with the help of our invented software. We carry out the injection of macroparticles into neon plasma using a cylindrical container with aluminum oxide particles which were situated in a volume of the discharge tube. The container construction and injection technology provided multiplication of particles, injected into plasma.

With the help of mentioned above equipment and methods of data acquisition we figured out the area existence of various phase conditions of plasma structure for the concrete "plasma-CDP" system (the neon glow discharge plasma and aluminum oxide particles). (fig. 4).

## 4. The investigation of ordered structure characteristics (the neon glow discharge plasma and aluminum oxide particles. Mean diameter is 30 $\mu$.).

The analysis of spatial-time distribution of macroparticles in a laser "knife" cross-section permit to set up shape, geometric size, ordering level, a number of macroparticles in the structure and an interpartical distance under various conditions (discharge current, pressure) in the glow discharge in the cylindrical tube with an internal diameter of d=27mm.

The truncated cone inside the tube was used to create a trap and localize ordered structure.

Experimental investigations were randomized. The discharge tube was degassed before each experiment. Then defined value of pressure was established and the macroparticles were injected into plasma by shaking the container a certain number of times.

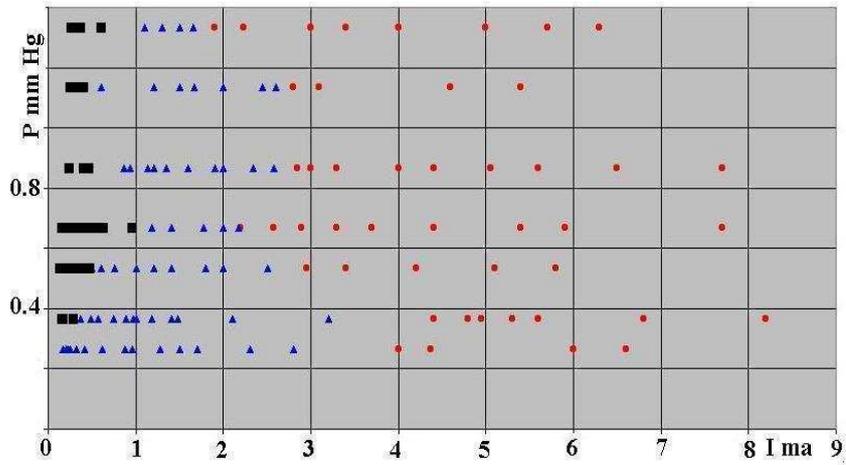

Fig.4. The area of existence of dusty structure of aluminum oxide particles in neon.
Black rectangles-"crystal", blue triangles -"liquid", red circles- "gas".

The structures, formed in the mentioned above conditions are presented in fig 5. for certain conditions (P=80Pa, i=0.3, 2мА, number of injections are 5<=N<=95).

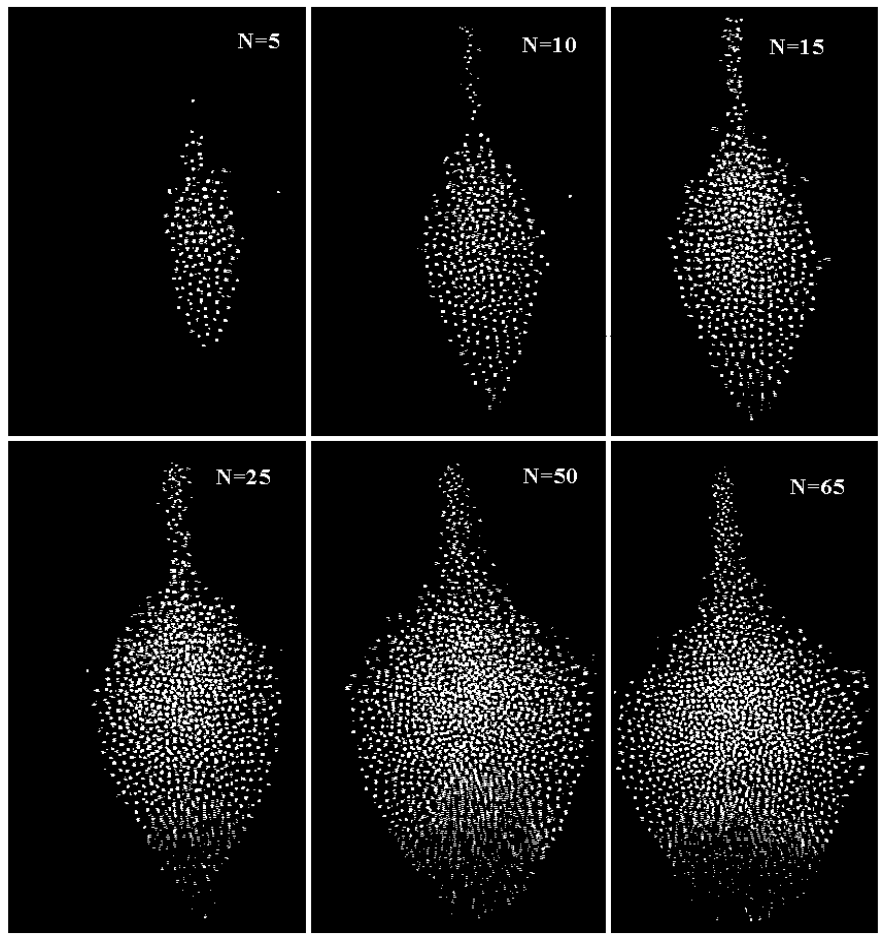

Fig. 5. The growth of ordered dusty structure.

The dependence of structure volume and number of particles in the structure on the number of injections one can see in fig. 6 and fig.7.

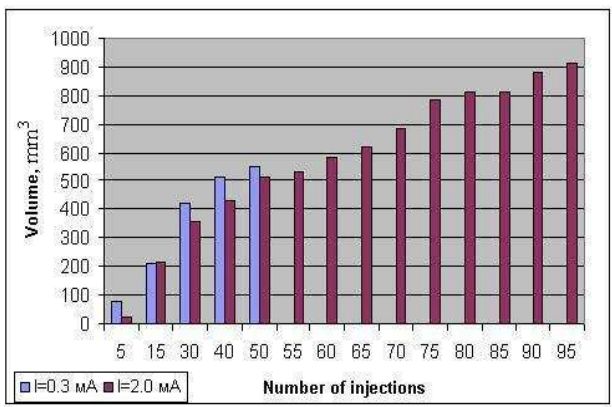
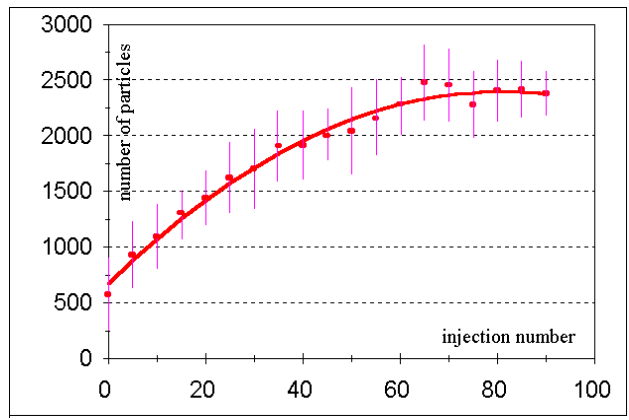

Fig. 6. The dependence of structure volume on the number of injections

Fig. 7. The dependence of particle number in the image of dusty cloud on the number of injections

Fig. 7 demonstrates the result of data averaging over the experiments which were carried out on different days according to mentioned above procedure.

Having analyzed the experimental results, we discovered the independence of distance between macroparticles from volume of dusty structure and value of interpartical distance is equal to **d=(134±7)** µ. This result was obtained in dusty structures, with a usage of 10 up to 95 injections. (The graph of experimental data of interpartical distances is presented in fig. 8.)

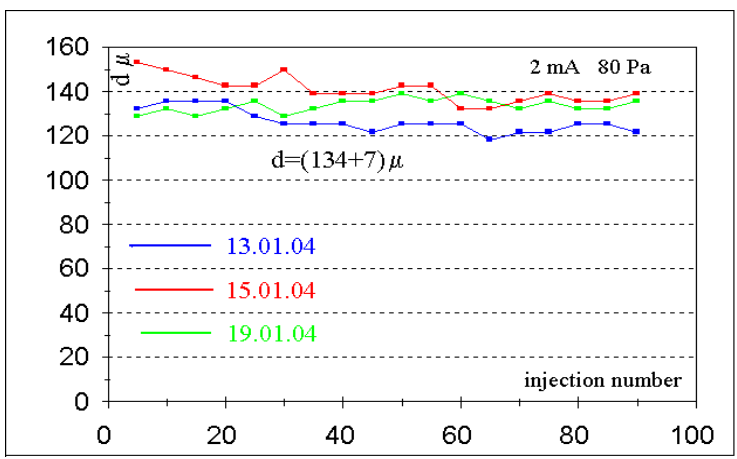

The analysis of the experimental data, presented in fig.5, enabled us to assume that the certain number of electrons and macroparticles in localization area is crucial for dusty structure forming. The scarcity of either of these components of complex plasma will limit structure volume (size).

Fig. 8. The interpartical distances as a function of injection number received on different days.

### 5. The influence of discharge current on ordered structure characteristics.
The described above characteristics of ordered dusty structure, grown to a certain size were investigated by discharge current changing under constant pressure.

This process is shown in pictures below (fig.9):

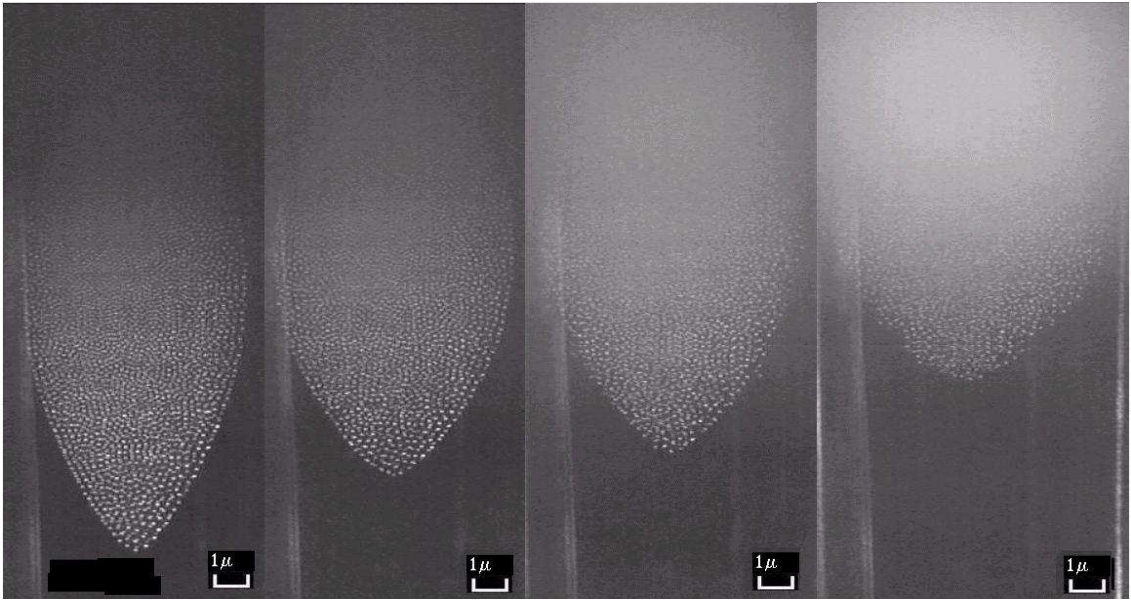

Fig. 9. Structure form changing by current heating I=0.3, 1, 3, 5мА.

The usage of the mentioned above methods of video processing allows to establish invariability of the structure volume (within the range of experimental errors) when discharge current rises up to 20 times, i.e. in the case of essential heat release into the structure volume (fig.10).

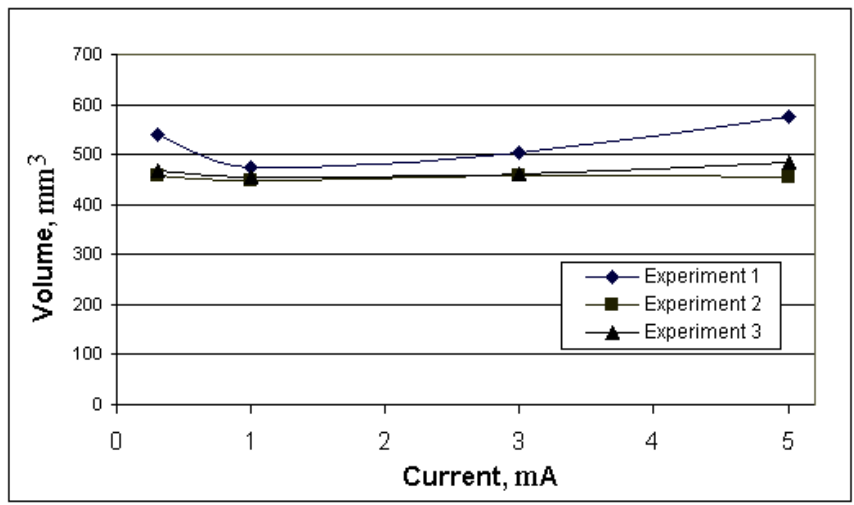

Fig.10. The dependence of structure volume on the discharge current.

However, at the same time crucial decrease of strength order due to particle oscillations' amplitude rising near the equilibrium position were observed. But mean interpartical distance remains statistically constant.

These facts permit to draw an analogy with the matter [4], which has phase transitions and the following characteristics: heat capacity, specific heat of fusion and others thermodynamic characteristics. We assume that the same numerical characteristics of our matter will be obtained.

Hence, we can draw a conclusion that there are certain types of matter that can exist as a plasma-CDP-force-field system. The role of the latter component should be separately analyzed.

**6. Conclusion.**

The experimental results, including the observation of shape and volume of ordered macroparticle structures, self-organizing in certain plasma conditions showed that:
1. the shape, size and concentration of CDP of macroparticles are defined by force fields of plasma and by plasma characteristics, as well as particles' size and matter.
2. a macroparticle-plasma system has properties similar to properties of matter which has different phase conditions and so such system is similar to certain matter.

Dusty structures as a kind of matter have not been adequately investigated. Additional complex investigations of dusty plasma will help to work out numerous applications of dusty structure.

**Acknowledgments**

The authors thank Prof. L.A. Luizova, Prof. V.I. Sysun (Petrozavodsk State University, Russia); Candidates of Science L.V. Deputatova, V.I. Molotkov and also V.I. Vladimirov, Doctor of Science V.C. Filinov (IHED RAS, Moscow, Russia) for being permanently interested and helpful; Prof. G. Morfill and Candidates of Science S.A. Khrapak (Max Planck Institute for Extraterrestrial Physics, Garching, Germany) for collaboration; engeneer A.I. Scherbina, students and postgraduates of REC "Plasma" of PetrSU for their assistance in experimental work.

This work was supported by grant PZ-013-02 CRDF, the Ministry of Education of the Russian Federation, the Government of Karelia, INTAS 2000-0522 grant and a state contract with the Ministry of Industry, Technology and Science of the Russian Federation.